\documentstyle[psfig]{EuroPhys}
\input EuroMacr.tex
\mathchardef\gacc="328

\begin{document}


\shorttitle{Fractal Growth from Local Instabilities}
\title{Fractal Growth from Local Instabilities}
\author{Raffaele Cafiero$^1$ and Guido Caldarelli$^2$}
\institute{$^1$ PMMH Ecole Sup. de Physique et de Chimie Industrielles
(ESPCI), 10, rue Vauquelin-75231 Paris Cedex 05 France \\
$^2$ Sezione INFM di Roma1 and Dipartimento di Fisica, 
Universit\`a "La Sapienza", P.le A.Moro 2 00185 Roma, Italy.}

\rec{ }{ }
\pacs{
\Pacs{61}{43.Hv}{Fractals; macroscopic aggregates 
(including diffusion-limited aggregates)}
\Pacs{68}{35.Ja}{Surface and interface dynamics and vibrations}
\Pacs{81}{10.Bk}{Growth from vapor}
\Pacs{64}{60.Ht}{Dynamic critical phenomena}
}

\maketitle

\begin{abstract}
We study, both with numerical simulations 
and theoretical methods, a cellular automata model
for continuum equations describing growth processes in the
presence of an external flux of particles. As a result of local
instabilities we find a fractal regime of growth for small
external fluxes. 
The growing tip is selected with probability proportional to the
curvature in the point.
A parameter $p$ gives the probability of lateral growth on the tip.
The value of $p$ determines the fractal dimension of the aggregate.
Furthermore, for each value of $p$ a cross-over between two different 
fractal dimensions is observed. Instead, the roughness exponent $\chi$ of the 
aggregates does not depend on $p$ ($\chi \simeq 0.5$). 
Fixed scale transformation approach is applied to compute theoretically 
the fractal dimension for one of the branches of the structure. 
\end{abstract}

From a technological point of view the growth by vapor atom deposition
is the most common way to produce films \cite{uno}.
It is known that local instabilities play
a decisive role in the formation of patterns observed in experiments,
such as grooves, mounds and metallurgical junctions \cite{due}.
These kinds of phenomena are usually described by the following 
continuum equation \cite{quattro}
\begin{equation}
\dot{h(x)} = J - \xi \nabla^2 h(x),
\label{piccoli gradienti}
\end{equation}
where $h(x)$ represents the surface height corresponding to
the substrate coordinate $x$; $J$ is the incident flux of particles.
The Laplacian term with the negative sign plays the role
of an antidiffusion, and it is responsible for the instability.
The precise meaning of $\xi$ depends on the particular kind of
instability involved in the process.
The $J$ term in eq.(\ref{piccoli gradienti}) can be ruled out with a change
of coordinates in a frame comoving with the surface. 
Then $h$ can be considered just as the fluctuation around the average.
The resulting equation is 
\begin{equation}
\dot{h}(\vec{x},t) = -\nabla^2h(\vec{x},t).
\label{surg}
\end{equation}
The solution of (\ref{surg}) in Fourier space is
\begin{equation}
h(\vec{k},t) = h(\vec{k},0)e^{\xi \vec{k}^2t}.
\end{equation}
Therefore all the modes present at the origin are exponentially 
amplified. In order to control this instability, people usually 
introduce a term proportional to $\nabla^4 h(x)$ in the eq. 
(\ref{piccoli gradienti}). This brings to
\begin{equation}
\dot{h}(\vec{k},t) = \xi \vec{k}^2h(\vec{k},t) -\lambda \vec{k}^4
h(\vec{k},t)
\end{equation}
whose solution now is
\begin{equation}
h(\vec{k},t) = h(\vec{k},0)e^{\xi \vec{k}^2-\lambda \vec{k}^4}t.
\end{equation}
Therefore only the long wavelength modes are still unstable.
The result of this process is the formation of columns
separated by very deep grooves.

An alternative continuum approach is represented by the
Reparametrization Invariant Formulation of the surface growth equation 
\cite{repinv}.
The basic idea is that the growth of surfaces only depends
on its intrinsic properties. Therefore eq.(\ref{piccoli gradienti}) can
be rewritten as
\begin{equation}
\dot{\vec{r}} = \hat{n} (J - \xi H)
\label{RIF}
\end{equation}
where $\vec{r}$ is the position vector of the surface, $\hat{n}$
the local normal unit vector and $H$ the local curvature. $J$ and $\xi$ are 
defined as before. Within this scheme the $\nabla^4 h$ term
is substituted by the Beltrami-Laplace operator for the curvature.

When the tips of two
columns join together the deep groove below them disappears and the
effective resulting surface is almost flat\cite{delose}.
In this spirit, a simpler way to approach the study of surface growth is
represented by the introduction of lattice models  
able to keep the relevant properties of the phenomenon. 
Many models have been proposed in the past years: the Eden model
\cite{eden}, 
ballistic deposition models\cite{ball}, Solid-On-Solid (SOS) models
\cite{solid} and others (for a review on surface growth models see 
also Ref.\cite{zhang,stanley}).

Here, we present a lattice model to verify the role of the surface 
curvature in these growth processes.
Consider a surface subject to an incoming flux of particles.
The model is based on the following two rules. 
Firstly, the probability that a particle stops in a given site
is proportional to the local negative curvature of the surface.
For each point the curvature is computed in a first
approximation as the Laplacian function in the point.
In this way, the growth probability at $t$ on the tip $\vec{x},t$ is 
given by
\begin{equation}
P(x,t) \propto [2 h(x,t)-h(x-1,t)-h(x+1,t)]
\end{equation}

Secondly,  a particle sticks right on the tip with probability $p$ while 
with probability $1-p$ it can stick laterally.
We introduce in this way the possibility to develop overhangs. 
Under this respect this model can be considered a generalization of the
growth model proposed in Ref.\cite{stick} for sticky particles.
The surface height $h(i)$ on a site $i$ corresponds to the position
of the highest particle on that column.
We will consider here the limit of vanishing external flux, which is
mimicked allowing just one particle to stick at a time.

This model of aggregation does not give rise to compact structures, but 
instead to fractal ones merging themselves to form a continuous surface.
In the system, every now and then, a tip whose value of the Laplacian
operator becomes larger and larger, spans uniformly the system size.
When the growth reaches a boundary there is a new flat surface 
where the deposition proceeds. 

\begin{figure}
\centerline{\psfig{figure=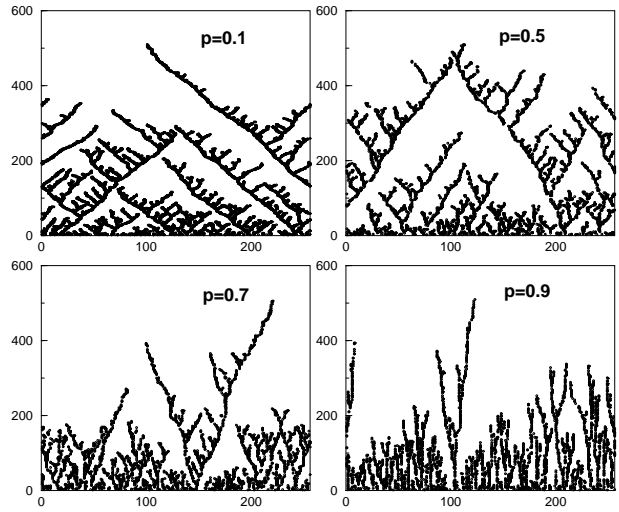,height=6.7cm}}
\caption{(a)-(d): Typical patterns for different values of $p$ 
(the system size and the values of $p$ are shown in the legends).}
\label{fig1}
\end{figure}

To reduce finite size effects, in our simulations we considered periodic 
boundary conditions along the horizontal direction of the system. 
In Fig.\ref{fig1} we present some typical patterns for size $L\times2L$ 
($L=256$) and for different values of $p$. There is a transient regime, 
that we call the {\em scaling regime } in analogy with 
the dielectric breakdown model (DBM) 
\cite{dbm}. Here several branches compete with each other in the 
growth process. After that, in the {\em steady state}, a 
single branch prevails 
and grows forever. 
Due to the screening induced by the shadowing between 
competing branches, the system develops fractal properties 
in the scaling regime where we focus our studies. 
We have performed several numerical 
simulations of this model for different values of $p$ 
and for system sizes up to $L=512$, and height $2L$. The results 
presented here refer to average on about $100$ realization for
every value of $p$ and $L=256,512$.

In this model the sample grows as a compact set of different columns 
as observed 
experimentally in \cite{colu1,colu}. Authors, in \cite{colu},  conjecture 
an exponential distribution for the tip height
from their experimental results on columnar structured gold 
electrodeposits. 
\begin{equation}
D(h)\propto exp(-bh)
\label{colgr}
\end{equation}
where $b$ is the inverse of a characteristic length. The same shape of $D(h)$ 
was previously observed in \cite{colu1}, through  scanning tunneling 
microscopy analysis of vapor deposited gold films. We computed $D(h)$ for our 
model with $p=1$, $L=256$ and maximum tip height $h_{max}=256,512$. The 
results reproduce the same exponential behavior 
of $D(h)$ observed in \cite{colu1,colu}. The characteristic height $b^{-1}$ 
turns out to be actually proportional to the maximum height $h_{max}$ 
($b^{-1}= c^{-1} h_{max} \sim \frac{h_{max}}{5}$).

\begin{figure}
\centerline{\psfig{figure=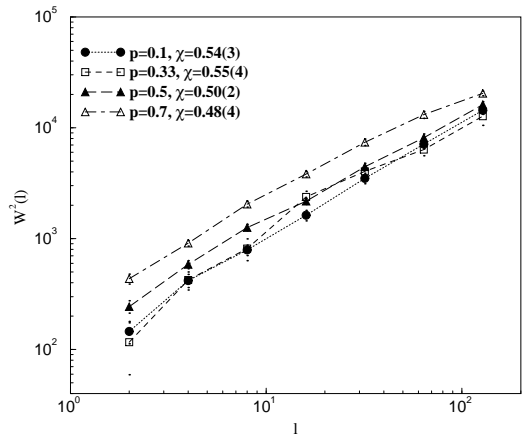,height=6.7cm}}
\caption{Scaling behavior of the squared width $W^2(l)$ width of portions 
of size $l$ of the aggregates surface, defined as the path of the highest 
occupied sites for each column, for different $p$. We get a roughness exponent $\chi$ 
around $1/2$ for all values of $p$.}
\label{fig3}
\end{figure}

In order to describe the clusters obtained we also focused on  
the roughness exponent $\chi (p)$ of the surface. 
The roughness exponent is defined as the ensemble 
averaged width of the interface
\begin{equation}
 W(l,t)= \langle(h(i,t)-\langle h(i,t)\rangle)^2\rangle^{1/2} 
\sim l^{\chi} f(t^{1/z}/l) 
\label{roughness}
\end{equation}
where $z$ is the dynamical exponent, the angular brackets
denote the average over all portions of the interface of length $l$ 
for a given realization of total length $L$, and over different realizations,
 and $f$ is a scaling
function such that $f(y) \sim y^{\chi}$  for $y<<1$ and $f(y)=const$
for $y>>1$. The exponent $\beta=\chi/z$ describes the transient
roughening. In the asymptotic time limit on gets:
\begin{equation}
W(l,t=\infty)=W(l)\propto l^{\chi}
\label{wasint}
\end{equation}
Interestingly enough, the value of this quantity does not depend on the value
of $p$ as one can see from Table \ref{tab1}. A plot of this scaling is
shown in Fig.\ref{fig3}. 

\begin{figure}
\centerline{\psfig{figure=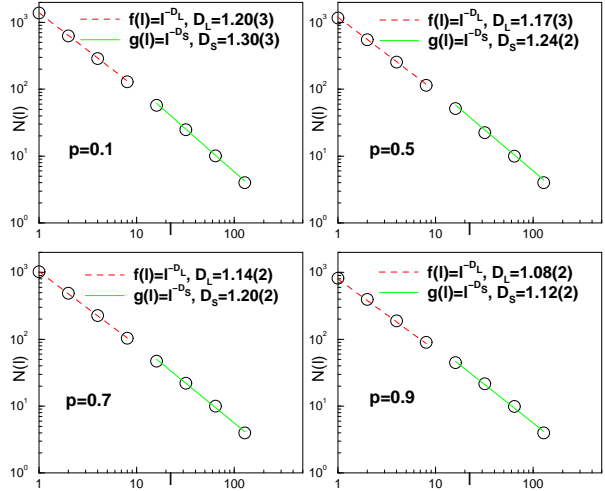,height=6.7cm}}
\caption{(a)-(d): Scaling behavior of the number of occupied boxes $N(l)$ 
of size $l$ versus $l$, for different values of $p$, 
horizontal size $L=256,512$ and height $h=2L$. 
The box counting is made on a strip of height $h=L/2$.}
\label{fig2}
\end{figure}

A dependence on $p$ affects, instead, the fractal dimension of the
aggregate. In our simulations we studied the lower strip 
where the growth process can be considered fully screened 
by the shadowing effect.
In Fig.\ref{fig2} we show the plot of $N(l)$, number of occupied boxes 
of size $l$, versus $l$ in the scaling regime for different values of $p$.
In Tab.\ref{tab1} we report the measured values for these exponents. 
The result is that the value of $D_f$ strongly depends on the parameter $p$.
Moreover, for all values of $p$ a cross-over between two different fractal 
dimensions is observed. 
In particular the region of small $l$ are characterized by a value of 
$D_f^{low}$ changing from $1.20(3)$ to $1.08(3)$ as $p$ passes from $0.1$ to $0.9$.
Larger regions are characterized by a similar trend with the value of 
$D_f^{high}$ changing from $1.30(3)$ to $1.12(3)$ as $p$ passes from $0.1$ to $0.9$

As regards the trend of this variation one can conclude that in the limit
$p\rightarrow 1$ the growth is dominated by the highest tip that grows forever. 
In this condition the growth of the aggregate  leaves large empty regions that
strongly reduce the fractal dimension of the aggregate.
The possibility of lateral growth instead, delays this effect in such a way to form
an intermediate region where one can study non trivial fractal properties.

\begin{figure}
\centerline{\psfig{figure=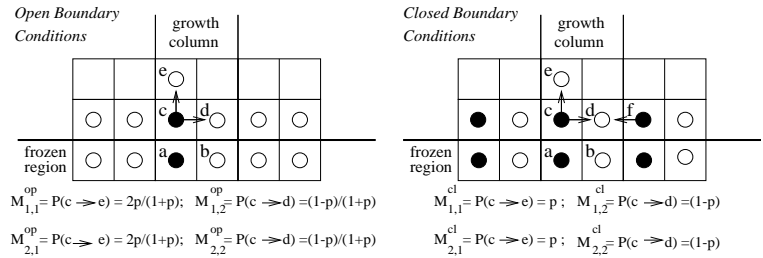,width=10cm}}
\vspace{0.5cm}
\caption{Lattice path integrals for the two types of boundary conditions 
considered. Please note that both in OBC than in the CBC, all
the processes starting from configuration of the type $C_2$ gives the same
results of processes starting from configuration $C_1$. In fact, if one 
consider the site b black, it is impossible to have 
growth process starting from it,
since the Laplacian operator in the point is negative}
\label{fig4}
\end{figure}

In order to obtain an analytical computation to reproduce the behavior of 
$D_f^{low}$, we used a theoretical scheme
of a novel type that can deal with irreversible dynamic processes
in a reasonably systematic way.
The Fixed Scale Transformation (FST) \cite{fst} focuses on the dynamics
at a given scale and allows to compute the nearest neighbors correlations
at this scale by lattice diagrams.
The structure that we want to study becomes well defined 
in the following limits:
\begin{itemize}
\item{asymptotic time limit ($t \rightarrow \infty$): only regions
of the system where the growth is frozen are considered;}
\item{large scale limit ($r \rightarrow \infty$) to avoid
the presence of finite size effects}
\end{itemize}
In the FST scheme, these two limits are reached through the
fixed point convergence of the lattice diagrams after a certain order
and considering the system at a coarse grained scale \cite{raf}.
In order to perform the first limit one defines an elementary cell
of two sites. The two configurations in Fig.\ref{fig4} 
are the only ones subject to growth, with an associated probability
to have a transition from one to the other in the growth
direction. 
The fixed point of the process (that can be regarded as the limit 
$t \rightarrow \infty$)
is obtained looking at the eigenvector corresponding to the eigenvalue $1$
(which is also the largest one)
\begin{equation}
M 
\left( \begin{array}{c}
C_1 \\ C_2
\end{array} \right)^t
=
\left( \begin{array}{c}
C_1 \\ C_2
\end{array} \right)^{t+1}
\label{FST}
\end{equation}
where $C_1$ and $C_2$ are the probabilities of occurrence
of configurations of the type $1$ and $2$ respectively.

The average number of boxes occupied by the structure in a change of 
scale $l \rightarrow \frac{l}{2}$ is
\begin{equation}
<n> = C_1 + 2 C_2.
\label{occupied boxes}
\end{equation}
Since we are studying the intersection set between the
structure and a line defining the growth front, the
fractal dimension of the structure is simply
\begin{equation}
D_f = 1 + D_t
\label{fractal dimension}
\end{equation}
where $D_t$ is the transverse fractal dimension related 
to $<n>$ by the formula:
\begin{equation}
D_t = \frac{\ln{<n>}}{\ln{(2)}}=\frac{\ln{(C_1 + 2 C_2)}}{\ln{(2)}}.
\label{transverse fractal dimension}
\end{equation}

Given this theoretical framework, it is very easy to develop the calculations 
for the present case.
Consider the graph in Fig.\ref{fig4}; as a first approximation
of the boundary conditions, one can consider the open ones, where the pattern
has no other occupied sites around.
The opposite condition is  represented by the closed ones
in which the entire pattern is replied. In both cases the growth probabilities 
are normalized to one inside the growth column.
Usually, a satisfactory approximation consists in considering 
only these two limiting cases, where the two fixed value of the configuration 
probabilities are given by:
\begin{eqnarray}
C^*_1\!&\!=\!&\!\frac{M_{12}^{cl}\!+\!2M_{21}^{cl}\!-\!\frac 3 2 M^{op}_{21}\!-\!
\!\sqrt{(\frac 3 2 M^{op}_{21}\!-\!M_{12}^{cl}-2M_{21}^{cl})^2\!
-\!4AM_{21}^{cl}} } {2A}  \nonumber \\
C^*_2\!&\!=\!&\!1-C^*_1 
\end{eqnarray}
where $A=M_{12}^{cl}+M_{21}^{cl}
-3/2(M^{op}_{12}+M^{op}_{21})$.
By using the scheme shown in Fig.\ref{fig4} one obtains the values of the
matrix elements in the FST scheme for open and closed boundary conditions.
\begin{eqnarray}
M_{12}^{op} = \frac{1-p}{1+p}&;&\mbox{   }M_{11}^{op} = 1-M_{12}^{op} \\
M_{21}^{op} = \frac{2p}{1+p}&;&\mbox{   }M_{22}^{op} = 1-M_{21}^{op} \\
M_{12}^{cl} = 1-p&;&\mbox{   }M_{11}^{cl} = 1-M_{12}^{cl}\\
M_{21}^{cl} = p&;&\mbox{   }M_{22}^{cl} = 1-M_{21}^{cl}
\end{eqnarray}
Two sources of screening are present in our model: one is due to 
the fact that only the tips can grow, and this is a local effect 
which is accounted for in the FST computation. The other, 
more important, is shadowing, which is highly non-local, depending 
on the whole structure of the aggregate. This effect cannot be 
easily accounted in the FST scheme \cite{fst}. This reflects on the results of 
the computation, especially for low values of $p$, where shadowing 
is more important. Instead for as $p \rightarrow 1$ there is a reasonable agreement 
between the FST computation and numerical simulations (last column of 
Table \ref{tab1}).

To conclude, we have presented a simple cellular automata model, 
which describes fractal surface growth in the limit of vanishingly 
small external flux, under the effect of a local instability. The model 
shows non-trivial, interesting fractal and roughness properties, which 
are studied both numerically and analytically.
We acknowledge useful discussion with R. Ball, P. De Los Rios, A. Maritan, 
L. Pietronero, and the support of the European Network contract FMRXCT980183.

\begin{table} 
\begin{centering} 
\protect \caption{Fractal dimension for the lower and upper scaling region 
and roughness exponent for different values of $p$ and system size $L=256,512$.} 
\label{tab1} 
\begin{tabular}{|c|c|c|c|c|} 
\hline 
$p$ & $D_f^{low}(p)$ & $D_f^{high}(p)$ & $\chi(p)$ & $D_f^{low}FST(p)$\\ 
\hline 
$0.1$ & $1.20\pm0.03$ & $1.30\pm0.03$ & $0.54\pm0.03$ & $1.91$ \\ 
$0.33$ & $1.18\pm0.03$ & $1.28\pm0.03$ & $0.55\pm0.04$ & $1.66$ \\ 
$0.5$ & $1.17\pm0.03$ & $1.24\pm0.02$ & $0.50\pm0.02$ & $1.46$ \\ 
$0.7$ & $1.14\pm0.02$ & $1.20\pm0.02$ & $0.48\pm0.04$ & $1.24$ \\ 
$0.9$ & $1.08\pm0.02$ & $1.12\pm0.02$ & $-$ & $1.08$ \\ 
\hline 
\end{tabular} 
\end{centering} 
\end{table}  

\end{document}